*Article*

# AI-Powered Camera and Sensors for the Rehabilitation Hand Exoskeleton

**Md Abdul Baset Sarker , Juan Pablo Sola-thomas, Masudul H. Imtiaz \***

Department of Electrical and Computer Engineering, Clarkson University, Potsdam, NY 13699, USA; sarkerm@clarkson.edu (M.A.B.S.), schumaj@clarkson.edu

\* Correspondence: mimtiaz@clarkson.edu

Due to Motor Neurone Diseases, a large population remains disabled worldwide, negatively impacting their independence and quality of life. This typically involves a weakness in the hand and forearm muscles, making it difficult to perform fine motor tasks such as writing, buttoning a shirt, or gripping objects. This project presents a vision-enabled rehabilitation hand exoskeleton to assist disabled persons in their hand movements. The design goal was to create an accessible tool to help with a simple interface requiring no training. This prototype is built on a commercially available glove where a camera and embedded processor were integrated to help open and close the hand, using air pressure, thus grabbing an object. An accelerometer is also implemented to detect the characteristic hand gesture to release the object when desired. This passive vision-based control differs from active EMG-based designs as it does not require individualized training. Continuing the research will reduce the cost, weight, and power consumption to facilitate mass implementation.

Keywords: Artificial intelligence (AI), Assistive technology, Computer vision, Deep learning, Gesture detection, Google Coral, Hand exoskeleton, Motor neuron disease, Vision-enabled rehabilitation.

---

## 1. Introduction:

Hand motor impairments due to conditions such as Motor Neurone Diseases (MND) significantly affect a person's ability to perform daily tasks, leading to a decreased quality of life and independence. Such impairments manifest as muscle weakness in the hand and forearm, making it difficult to perform fine motor tasks like writing, buttoning a shirt, or gripping objects. Traditional rehabilitation methods include task-oriented training and the use of assistive devices, but these approaches often require extensive training and are not universally accessible.

Recent advancements in rehabilitation technology have focused on enhancing therapy through the use of robotic exoskeletons. These devices, such as the Pediatric Whole Hand Exoskeleton (PEXO) [1], offer promising solutions by providing actuated assistance tailored to the needs of children with neuromotor disorders, improving their ability to engage in functional tasks. Advancements in wearable robotics have shown significant potential in enhancing the rehabilitation process. For instance, devices like the SCRIPT passive orthosis and the Gloreha hand robotic rehabilitation system have made strides in this field by focusing on hand and wrist rehabilitation through interactive and dynamic assistance [2,3]. Moreover, recent studies highlight the importance of integrating user-friendly and intuitive control systems to maximize the effectiveness of these rehabilitation tools [4–9]. However, many existing systems, including [10], rely on active control methods like electromyography (EMG), which require individualized training and adaptation.

To address these limitations, our research presents a vision-enabled rehabilitation hand exoskeleton designed to assist individuals with hand impairments through a simple, intuitive interface.

Unlike EMG-based systems, our prototype integrates a camera and embedded processor to automate hand movements using air pressure, allowing for object manipulation without the need for user training. Additionally, an accelerometer detects characteristic hand gestures to facilitate object release, enhancing the overall usability and accessibility of the device [11].

Given the widespread use of camera-based solutions across different systems [12,13], we have adopted this approach in our design. For projects focused on image analysis, single-board computers like the Jetson Nano [14–18], Google Coral [19,20], and Raspberry Pi [21–23] are favored for their compact size and low power consumption. To optimize our system, we have used the Google Coral Dev Board Mini to run image processing and control peripherals because of its smaller size and inclusion of a Tensor Processing Unit (TPU) for faster processing of deep learning models.

In our previous research, we implemented a vision-based system on a prosthetic hand [20]. In this project, we plan to implement a vision-based system for an exoskeleton. By employing vision-based control, this project aims to reduce rehabilitation exoskeletons' cost, weight, and power consumption, making them more feasible for widespread implementation. This approach will simplify the user interface as well as significantly improve the independence and quality of life for individuals with motor neuron syndrome and other hand motor impairments. Figure 1 shows the prototype of AI cameras and sensors for hand exoskeleton with battery.

The contributions of this project are,
1. We have implemented a novel vision-based system in an exoskeleton for hand
2. An object detection model was trained with a custom dataset collected by the developed system

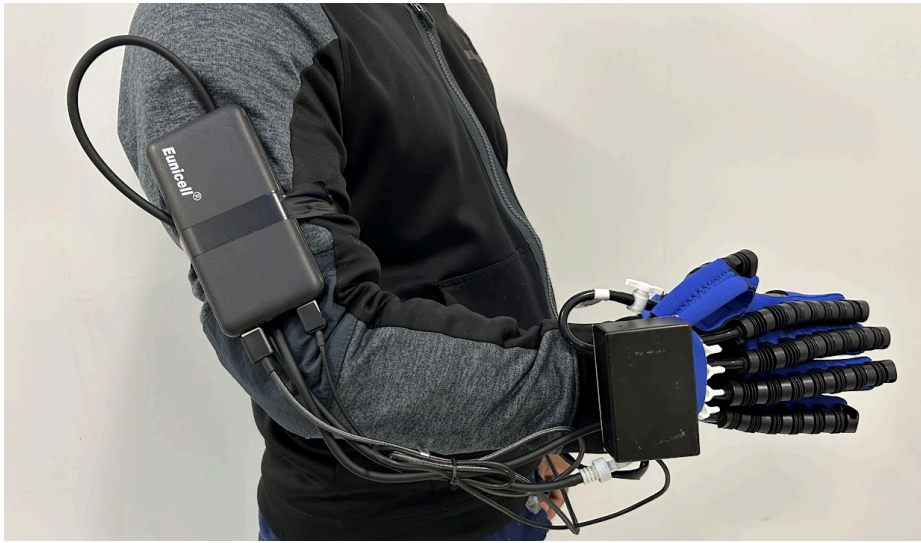

**Figure 1.** Prototype of AI cameras and sensors for hand exoskeleton

**2. Methodology:**

Figure 2 shows the block diagram of the vision-enabled rehabilitation hand exoskeleton. It shows the interconnected components that collectively enable automated hand movements. The system's core is a camera and embedded processor, Coral Dev Board mini [24]. The camera (5MP Coral camera [25]), Time of flight (TOF) distance sensor (VL6180X), and accelerometer are connected to the main processor. The camera captures real-time images, which the embedded processor. The switching circuit controls the air pressure system that drives the exoskeleton's finger movement through a solenoid, facilitating the opening and closing of the hand to grasp objects. Additionally, an accelerometer (ADXL345) is integrated to detect specific hand gestures, allowing for the release of objects.

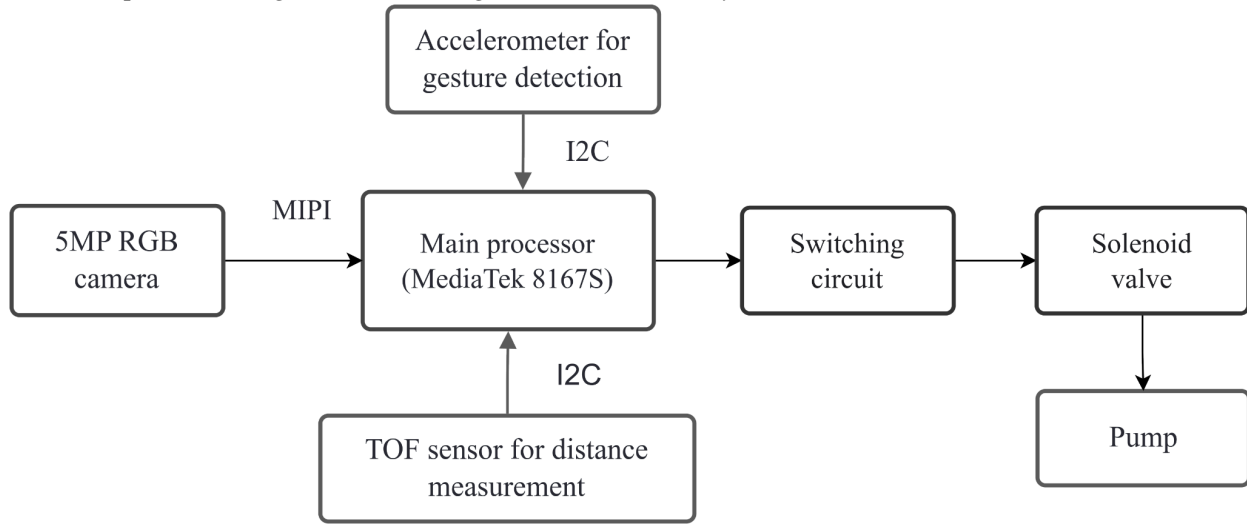

**Figure 2.** Block diagram of Exoskeleton

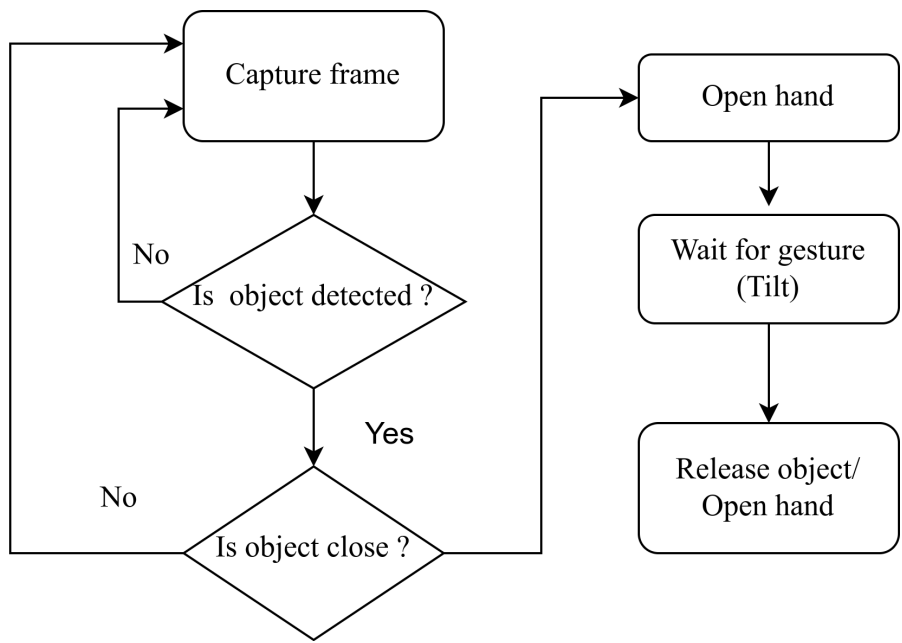

**Figure 3.** Flow chart of the system firmware

The flowchart of the vision-enabled rehabilitation hand exoskeleton is shown in Figure 3. The image capture stage involves the camera capturing real-time images of the user's hand and its environment. These images are then processed in the Object Detection phase, where computer vision algorithms analyze the visual data to identify the target object. Once the object is determined, the system moves to the control signal generation stage, where the control unit interprets the processed data and generates the appropriate control signals. These signals are then sent to the Air Pressure System, which adjusts the pressure in the actuators to mimic natural hand movements for grasping the object. During this process, the Gesture Detection stage continuously monitors the user's hand gestures via an accelerometer, allowing for intuitive commands such as releasing the object. The final Action Execution phase involves the actual movement of the exoskeleton's hand to grasp or release the object. Figure 4 shows a vertical and horizontal view of the vision-based exoskeleton and an example of detecting and grasping the object.

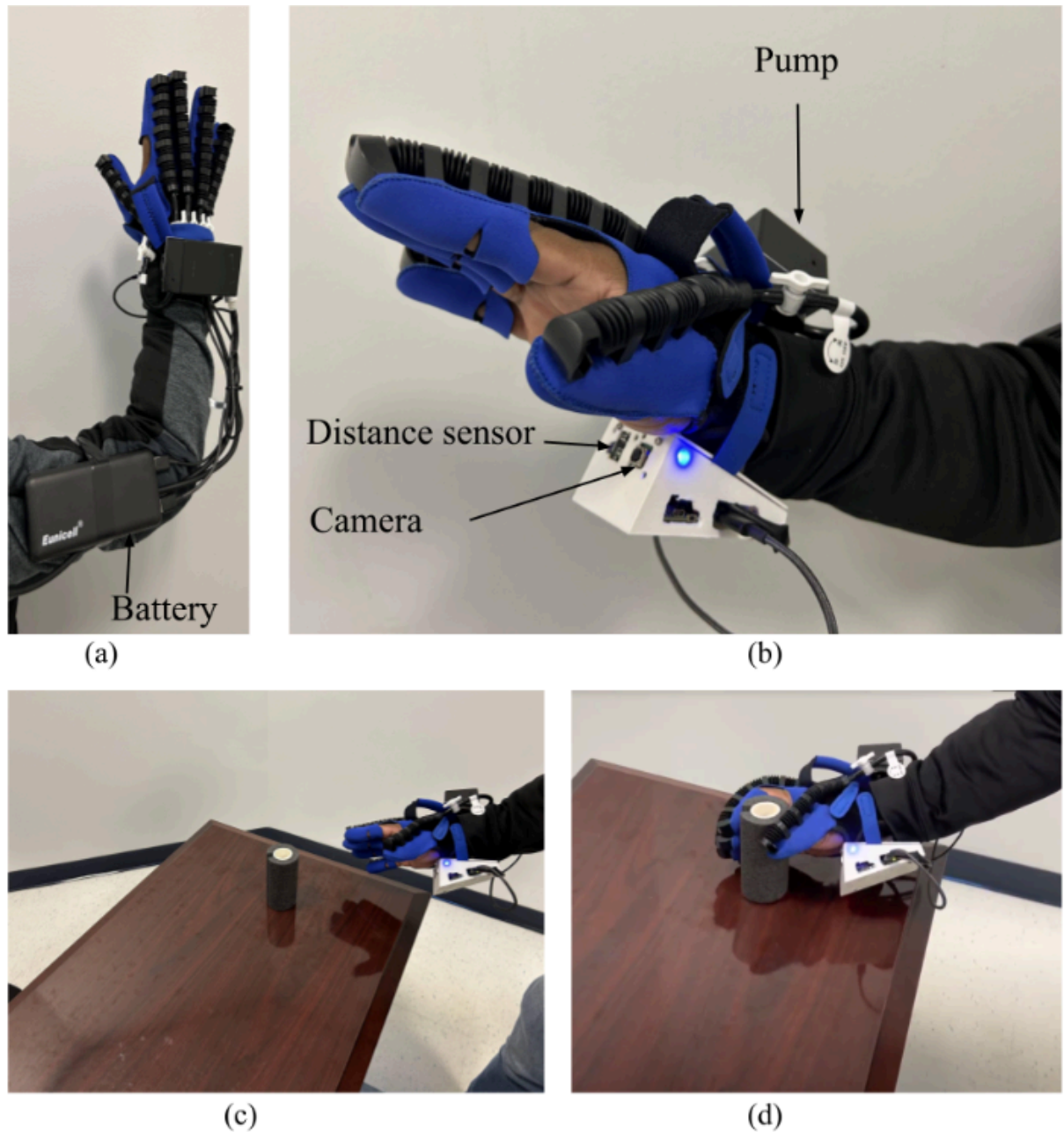

Figure 4. (a) Vertical and (b) horizontal view of the vision-based exoskeleton, (c, d) detecting and grasping the object.

*2.1. Hardware*

The main board we used for this exoskeleton is the Google Coral Dev Board mini [24]. The Coral Dev Board Mini is a compact single-board computer designed to make machine learning (ML) inferencing faster. While primarily intended as an evaluation device for the Accelerator Module, which includes the Edge TPU, it also functions as a fully operational embedded system suitable for a wide range of on-device ML projects. This board contains MediaTek 8167s System on Chip (SoC), a quad-core Arm Cortex-A35 processor designed for efficient performance and energy management. The IMG PowerVR GE8300 GPU is integrated within this SoC, ensuring reliable graphics processing. Google Edge TPU coprocessor delivers 4 TOPS (int8) and achieves an impressive efficiency of 2 Tera-Operations Per Second (TOPS) per watt. Complementing these components is 2 GB of LPDDR3 RAM and 8 GB of eMMC flash memory, providing adequate memory and storage capacity for various applications. We took the Hemiplegia Hand [26], which is for training and rehabilitation with preprogrammed functions, and replaced only the gloves with our control system.

This exoskeleton utilizes a 5-megapixel camera module compatible with Coral boards for capturing images [25]. The camera connects via the MIPI-CSI interface, and features autofocus, an 84/87-degree view angle, and various auto controls for image optimization. A VL6180X Time of Flight distance sensor is integrated for precise object distance detection. Additionally, an ADXL345 accelerometer detects gestures (tilt) to release objects.

*2.2. Training and Object Detection*

For object detection, we implemented the EfficientDet-Lite0 model architecture [27] as it is lightweight. For the training process, we gathered images of two objects and annotated all images using the open-source LabelImg script [28]. The images were divided into 10% for testing, 10% for validation, and 80% for training and trained for 300 epochs.

**3. Result and Discussion:**

The development of this novel vision-enabled rehabilitation hand exoskeleton represents a significant advancement in assistive technology for individuals with hand motor impairments. This system integrates a camera and embedded processor to enable automated hand movements, which simplifies the user interface and eliminates the need for extensive training typically required by EMG-based systems. By utilizing computer vision algorithms for object detection and an accelerometer for gesture recognition, the exoskeleton provides a more intuitive and accessible solution for users. One of this system's primary advantages is its potential to reduce costs, weight, and power consumption, making it feasible for mass production and widespread use. The focus on affordability and portability addresses some of the critical barriers to the adoption of advanced rehabilitation devices. Moreover, the use of air pressure actuators, controlled through vision-based algorithms, offers a natural and responsive user experience, mimicking the subtle movements of the human hand.

Despite these advancements, several challenges and limitations need to be addressed. The accuracy and speed of the object detection algorithms are crucial for real-time application, and any delays could impact the effectiveness of the device. Additionally, the robustness of the system in various lighting conditions and environments must be ensured to make it reliable for everyday use. Further research is needed to optimize the image processing algorithms and improve the gesture detection sensitivity of the accelerometer.

We have tested this device in the lab, from object detection to grasping and releasing. Object detection runs at six frames per second. The pilot testing phase demonstrated promising results, with users successfully performing grasp and release tasks. However, extensive clinical trials involving a larger and more diverse group of participants are necessary to validate the system's efficacy and usability comprehensively. User feedback during these trials will be invaluable for refining the design and functionality of the exoskeleton.

Compared to other existing solutions, such as the PEXO [1] and SCRIPT [2,29], EMG-based [10] systems, the vision-enabled exoskeleton offers a novel approach that simplifies user interaction while maintaining functional effectiveness. However, continuous improvements and iterations are essential to enhance its performance and user satisfaction.

**4. Conclusion**

The proposed vision-enabled rehabilitation hand exoskeleton presents a promising tool for assisting individuals with hand impairments, offering a blend of advanced technology and user-centered design. By addressing current limitations and incorporating user feedback, this system has the potential to significantly improve the quality of life and independence of its users. Future work will focus on refining the technology, conducting comprehensive clinical evaluations, and exploring additional applications and functionalities to broaden its impact.

**Reference**
1. Bützer, T.; Dittli, J.; Lieber, J.; van Hedel, H.J.A.; Meyer-Heim, A.; Lambercy, O.; Gassert, R. PEXO - A Pediatric Whole Hand Exoskeleton for Grasping Assistance in Task-Oriented Training. In Proceedings of the 2019 IEEE 16th International Conference on Rehabilitation Robotics (ICORR); June 2019; pp. 108–114.
2. Ates, S.; Haarman, C.J.W.; Stienen, A.H.A. SCRIPT Passive Orthosis: Design of Interactive Hand and Wrist Exoskeleton for Rehabilitation at Home after Stroke. *Auton. Robots* **2017**, *41*, 711–723, doi:10.1007/s10514-016-9589-6.
3. Gloreha—Hand Robotic Rehabilitation: Design, Mechanical Model, and Experiments | J. Dyn. Sys., Meas., Control. | ASME Digital Collection Available online: https://asmedigitalcollection.asme.org/dynamicsystems/article/138/11/111003/473979/Gloreha-Hand-Robotic-Rehabilitation-Design (accessed on 7 June 2024).
4. Polygerinos, P.; Wang, Z.; Galloway, K.C.; Wood, R.J.; Walsh, C.J. Soft Robotic Glove for Combined Assistance and At-Home Rehabilitation. *Robot. Auton. Syst.* **2015**, *73*, 135–143, doi:10.1016/j.robot.2014.08.014.
5. Hofmann, U.A.T.; Bützer, T.; Lambercy, O.; Gassert, R. Design and Evaluation of a Bowden-Cable-Based Remote Actuation System for Wearable Robotics. *IEEE Robot. Autom. Lett.* **2018**, *3*, 2101–2108, doi:10.1109/LRA.2018.2809625.
6. Nycz, C.J.; Meier, T.B.; Carvalho, P.; Meier, G.; Fischer, G.S. Design Criteria for Hand Exoskeletons: Measurement of Forces Needed to Assist Finger Extension in Traumatic Brain Injury Patients. *IEEE Robot. Autom. Lett.* **2018**, *3*, 3285–3292, doi:10.1109/LRA.2018.2852769.
7. Arata, J.; Ohmoto, K.; Gassert, R.; Lambercy, O.; Fujimoto, H.; Wada, I. A New Hand Exoskeleton Device for Rehabilitation Using a Three-Layered Sliding Spring Mechanism. In Proceedings of the 2013 IEEE International Conference on Robotics and Automation; May 2013; pp. 3902–3907.
8. Aubin, P.; Petersen, K.; Sallum, H.; Walsh, C.; Correia, A.; Stirling, L. A Pediatric Robotic Thumb Exoskeleton for At-Home Rehabilitation : The Isolated Orthosis for Thumb Actuation (IOTA). *Int. J. Intell. Comput. Cybern.* **2014**, *7*, 233–252, doi:10.1108/IJICC-10-2013-0043.
9. Bianchi, M.; Secciani, N.; Ridolfi, A.; Vannetti, F.; Pasquini, G. Kinematics-Based Strategy for the Design of a Pediatric Hand Exoskeleton Prototype. In Proceedings of the Advances in Italian Mechanism Science; Carbone, G., Gasparetto, A., Eds.; Springer International Publishing: Cham, 2019; pp. 501–508.
10. Ryser, F.; Bützer, T.; Held, J.P.; Lambercy, O.; Gassert, R. Fully Embedded Myoelectric Control for a Wearable Robotic Hand Orthosis. In Proceedings of the 2017 International Conference on Rehabilitation Robotics (ICORR); July 2017; pp. 615–621.
11. Wu, J.; Pan, G.; Zhang, D.; Qi, G.; Li, S. Gesture Recognition with a 3-D Accelerometer. In Proceedings of the Ubiquitous Intelligence and Computing; Zhang, D., Portmann, M., Tan, A.-H., Indulska, J., Eds.; Springer: Berlin, Heidelberg, 2009; pp. 25–38.
12. Zhao, X.; Chen, W.-H.; Li, B.; Wu, X.; Wang, J. An Adaptive Stair-Ascending Gait Generation

Approach Based on Depth Camera for Lower Limb Exoskeleton. *Rev. Sci. Instrum.* **2019**, *90*, 125112, doi:10.1063/1.5109741.
13. Liu, D.-X.; Xu, J.; Chen, C.; Long, X.; Tao, D.; Wu, X. Vision-Assisted Autonomous Lower-Limb Exoskeleton Robot. *IEEE Trans. Syst. Man Cybern. Syst.* **2021**, *51*, 3759–3770, doi:10.1109/TSMC.2019.2932892.
14. Sarker, M.A.B.; Hossain, S.M.S.; Venkataswamy, N.G.; Schuckers, S.; Imtiaz, M.H. An Open-Source Face-Aware Capture System. *Electronics* **2024**, *13*, 1178, doi:10.3390/electronics13071178.
15. salih, T.A.; Gh, M.B. A Novel Face Recognition System Based on Jetson Nano Developer Kit. *IOP Conf. Ser. Mater. Sci. Eng.* **2020**, *928*, 032051, doi:10.1088/1757-899X/928/3/032051.
16. Sarker, M.A.B.; Sola, E.S.; Jamieson, C.; Imtiaz, M. Autonomous Movement of Wheelchair by Cameras and YOLOv7. In Proceedings of the 3rd International Electronic Conference on Applied Sciences; December 2022.
17. Sola-Thomas, E.; Baser Sarker, M.A.; Caracciolo, M.V.; Casciotti, O.; Lloyd, C.D.; Imtiaz, M.H. Design of a Low-Cost, Lightweight Smart Wheelchair. In Proceedings of the 2021 IEEE Microelectronics Design Test Symposium (MDTS); May 2021; pp. 1–7.
18. Caracciolo, M.V.; Casciotti, O.; Lloyd, C.D.; Sola-Thomas, E.; Weaver, M.; Bielby, K.; Sarker, M.A.B.; Imtiaz, M.H. Autonomous Navigation System from Simultaneous Localization and Mapping. In Proceedings of the 2022 IEEE Microelectronics Design Test Symposium (MDTS); 2022.
19. Winzig, J.; Almanza, J.C.A.; Mendoza, M.G.; Schumann, T. Edge AI - Use Case on Google Coral Dev Board Mini. In Proceedings of the 2022 IET International Conference on Engineering Technologies and Applications (IET-ICETA); October 2022; pp. 1–2.
20. Sarker, M.A.B.; Sola, J.P.; Jones, A.; Laing, E.; Sola, E.S.; Imtiaz, M.H. Vision Controlled Sensorized Prosthetic Hand. In Proceedings of the Interdisciplinary Conference on Mechanics, Computers and Electronics (ICMECE); 2022.
21. Burns, C. Myo Gesture Control Band Controls MPL Prosthetic Arm Available online: https://www.slashgear.com/myo-gesture-control-band-controls-mpl-prosthetic-arm-18423486/ (accessed on 3 April 2023).
22. Joe, H.; Kim, H.; Lee, S.-J.; Park, T.S.; Shin, M.-J.; Hooman, L.; Yoon, D.; Kim, W. Factors Affecting Real-Time Evaluation of Muscle Function in Smart Rehab Systems. *ETRI J.* **2023**, *45*, 603–614, doi:10.4218/etrij.2021-0417.
23. Sarker, M.A.B.; Imtiaz, M.H.; Mamun, S.M.M.A. Development Of A Raspberry Pi Based Home Automation System. *Bangladesh J. Phys.* **2014**, *16*, 59–66.
24. Dev Board Mini Available online: https://coral.ai/products/dev-board-mini/ (accessed on 19 May 2022).
25. Camera Available online: https://coral.ai/products/camera (accessed on 20 May 2022).
26. Amazon.Com: Rehabilitation Robot Gloves Upgrade Hemiplegia Hand Stroke Recovery Equipment with USB Chargeable and Strength Adjustment : Health & Household Available online: https://www.amazon.com/Rehabilitation-Hemiplegia-Equipment-Chargeable-Adjustment/dp/B0BHNY5YF3/ref=sr_1_7?crid=1RVXEXKCA3NCH&dib=eyJ2IjoiMSJ9.au3PPA8ME-qYw4z7sSLBlytdX6kLHUPsnpBBMzOGmVV0duAvuIJFDyhsJN8HFrsvUNa9qgz67kirMVIr6JQrnd8Q0JLEW03ftlGfXJGH2VVrtKPMCN1ZdvJfvADGw0_bNMT5PftXAJ_iF1lzPHL2Rp4wR7BhcHs9Zv6vpAqOHgglW_ftkpH8JFTuY8OA8K-_VQoaTggNA8z3wgjupsy4H9SosK1n3lwkHT5lasQ9MoeX-L5TBmOrTKgbWChU7IcuutC-rd1yt4QIO__Xw4fnVB5gh2Qm5Ld8Rs3vZhhSL6w.A6-0NXA8S6y-A2C4rlrdX4WpyjEEPgQe4En1p7Wi7-U&dib_tag=se&keywords=rehabilitation%2Bglove&qid=1717704544&sprefix=rehevilatation%2Bglove%2Caps%2C82&sr=8-7&th=1 (accessed on 6 June 2024).
27. Tan, M.; Pang, R.; Le, Q.V. EfficientDet: Scalable and Efficient Object Detection.; 2020; pp. 10781–10790.
28. Heartexlabs/labelImg 2022, https://github.com/HumanSignal/labelImg (accessed on 19 May 2022)

29. Fajardo, J.; Ferman, V.; Cardona, D.; Maldonado, G.; Lemus, A.; Rohmer, E. Galileo Hand: An Anthropomorphic and Affordable Upper-Limb Prosthesis. *IEEE Access* **2020**, *8*, 81365–81377, doi:10.1109/ACCESS.2020.2990881.